# Phase-field crystal model for ordered crystals

Eli Alster,[1,2] K. R. Elder,[3] Jeffrey J. Hoyt,[4] and Peter W. Voorhees[2,*]

[1]*Department of Chemical and Biological Engineering,*
*Northwestern University, Evanston, Illinois 60208, USA*
[2]*Department of Materials Science and Engineering,*
*Northwestern University, Evanston, Illinois 60208, USA*
[3]*Department of Physics, Oakland University, Rochester, Michigan 48309, USA*
[4]*Department of Materials Science and Engineering and Brockhouse Institute for Materials Research,*
*McMaster University, Hamilton, ON, Canada L8S-4L7*

We describe a general method to model multicomponent ordered crystals using the phase-field crystal (PFC) formalism. As a test case, a generic B2 compound is investigated. We are able to produce a line of either first-order or second-order order-disorder phase transitions, features that have not been incorporated in existing PFC approaches. Further, it is found that the only elastic constant for B2 that depends on ordering is $C_{11}$. This B2 model was then used to study antiphase boundaries (APBs). The APBs were shown to reproduce classical mean field results. Dynamical simulations of ordering across small-angle grain boundaries predict that dislocation cores pin the evolution of APBs.

## I. INTRODUCTION

A fundamental problem in science is understanding multiscale phenomena. As a result of decades of research, molecular dynamics (MD) is now powerful enough to accurately predict the structure of quickly folding proteins [1] and faceting at asymmetrical grain boundaries [2] with atomic resolution. However, MD is fundamentally constrained by the fastest vibrational frequency of the system, so time scales beyond a microsecond are normally inaccessible. Unfortunately, most interesting phenomena in materials science occur over much longer time scales.

On the other extreme, phase field (PF) methods are able to capture mesoscale phenomena. PF theory is based on a free energy of the system that is a functional of the various order parameters, statistically averaging of short time and length scales, and as a result permits simulation of much longer times and distances [3]. However, this spatial averaging results in the loss of atomic features such as grain boundary and dislocation structure.

The phase field crystal (PFC) method operates in between these two regimes. Like traditional PF models, PFC theory involves a free energy functional, and it averages over rapid fluctuations in time to give a time scale of evolution on the order of diffusion rather than atomic vibration. However, unlike traditional PF, the free energy does not average over atomic distances, resulting in pattern formation at equilibrium. The simplest such free energy was inspired by the Swift-Hohenberg equation [4],

$$F = \int_V \left\{ \frac{\phi}{2}\left[-\varepsilon + (1+\nabla^2)^2\right]\phi + \frac{\phi^4}{4}\right\} d\mathbf{r}, \quad (1)$$

where $\phi$ is interpreted as the nondimensional atomic density. Although originally phenomenologically motivated,

PFC can be derived as a crude approximation of classical density functional theory [5]. The simple PFC model in Eq. 1 gives accurate descriptions of elasticity: it reproduces both Read and Shockley grain boundary energies and Matthews and Blakeslee misfit dislocation behavior during epitaxy [4]. Further, this model with minor modifications has been quantitatively fit to BCC iron [6, 7].

Unfortunately, Eq. 1 can only produce stripes, rods, and BCC for small $\varepsilon$ [8]. Consequently, there has been significant effort to adapt the free energy to reproduce other types of crystal symmetries including FCC, HCP, simple cubic, diamond cubic, and honeycomb [9–13] and to model binary and ternary alloys [5, 14–17]. Models have also been created to include vapor phases [18, 19], and electromagnetic effects [20, 21].

Current models, however, are unable to describe any structures with sublattice ordering, even structures as simple as the B2 CsCl structure, despite these structures making up a significant fraction of actual compounds of experimental interest. Interesting B2 compounds include the highly ductile rare earth intermetalics (YAg, YCu, DyCu) [22] and metal aluminides with high temperature structural stability (FeAl, CoAl, NiAl) [23]. The vast majority of so-called two-dimensional materials beyond graphene such as BN and $MoS_2$ also all fall under the general category of ordered crystals [24]. In order to study chemical vapor deposition growth, grain structure, grain boundary mobility, and elastoplasticity for these compounds under the PFC methodology, existing models do not suffice.

Further, because nearly all existing PFC models focus on solid-liquid coexistence, the only example of a second-order transition line in PFC theory is Seymour et al.'s paramagnetic to ferromagnetic transition [21]. However, order-disordered transitions can be both first-order, for example $Cu_3Au$ ($L1_2$) and DyCu (B2) [25], or second-order, for example CuZn (B2) and $Fe_3Al$ ($D0_3$) [26].

---

[*] p-voorhees@northwestern.edu

Second-order transitions are interesting as not only do they exhibit their namesake discontinuity in the second derivative of the free energy, they also do not form wetted domain boundaries, but instead exhibit correlation lengths between antiphase boundaries (APBs) that diverge as the phase boundary is approached [27].

This paper develops an equimolar binary PFC model that allows for sublattice ordering and that can be used to model a wide class of compounds. Section II derives the model directly from classical density functional theory, describes the general procedure for modeling any compound of interest, and implements a more appropriate temperature dependence of the Debye-Waller factor than used in the original structural PFC (XPFC) models [9, 10]. Section III examines the analytical behavior, elasticity, and numerical phase diagram for the specific case of a B2 compound. Section IV applies this B2 model to the study of antiphase boundaries and their dynamical interactions with grain boundaries.

## II. ORDERED BINARY PFC MODEL

### A. Derivation from CDFT

Classical density-functional theory postulates that the free energy of a system is a functional of its density. If we expand the Helmholtz free energy around a constant state (i.e., liquid) to second order then

$$F[\rho_A(\mathbf{r})] \simeq F[\rho_{A0}] + \int_V \frac{\delta F[\rho_A]}{\delta \rho_A(\mathbf{r})}\bigg|_{\rho_A=\rho_{A0}} \Delta\rho_A(\mathbf{r})d\mathbf{r} + \frac{1}{2!}\int_V \int_{V_1} \frac{\delta^2 F[\rho_A]}{\delta \rho_A(\mathbf{r})\delta \rho_A(\mathbf{r}_1)}\bigg|_{\rho_A=\rho_{A0}} \Delta\rho_A(\mathbf{r})\Delta\rho_A(\mathbf{r}_1)d\mathbf{r}_1 d\mathbf{r}, \quad (2)$$

where $\rho_A$ is the density, $\rho_{A0}$ is the reference density, and $\Delta\rho_A = \rho_A - \rho_{A0}$. In all analysis that follows, constants in energy will be ignored because they do not affect behavior. The first variation in Eq. 2 does not depend on interactions and is consequently just the entropic ideal free energy. The second term has been identified with the isotropic two-particle direct correlation function by the classical density-functional theory of freezing [28]. With these changes, the modified one component free energy is

$$F[\rho_A] = \int_V \left( \rho_A \ln \frac{\rho_A}{\rho_{A0}} - \Delta\rho_A - \frac{1}{2}\Delta\rho_A C_{AA}^{(2)} * \Delta\rho_A \right) d\mathbf{r} \quad (3)$$

where $C_{AA}^{(2)} * \Delta\rho_A = \int_{V_1} C_{AA}^{(2)}(|\mathbf{r}_1 - \mathbf{r}|)\Delta\rho_A(\mathbf{r}_1)d\mathbf{r}_1$ and $C_{AA}^{(2)}$ is the two-particle correlation function. For the case of a two component system,

$$F = F_A + F_B - kT \int \Delta\rho_A C_{AB}^{(2)} * \Delta\rho_B d\mathbf{r} \quad (4)$$

since $C_{AB}^{(2)} = C_{BA}^{(2)}$, where $C_{AB}^{(2)}$ is a partial correlation function between A and B [5].

Performing the substitutions $\rho = \rho_A + \rho_B$, $c = \rho_A/\rho$, and $\rho_0 = \rho_{A0} + \rho_{B0}$ results in

$$F = kT \int \left\{ \rho \ln \frac{\rho}{\rho_0} - (\rho - \rho_0) - \frac{1}{2}\rho \Big[cC_{AA}^{(2)}c + (1-c)C_{BB}^{(2)}(1-c) + 2cC_{AB}^{(2)}(1-c)\Big]\rho + \rho\Big[(1-c)\ln(1-c) + c\ln c\Big] + \rho c \Big[(C_{AA}^{(2)} - C_{AB}^{(2)})\rho_{A0} + (C_{AB}^{(2)} - C_{BB}^{(2)})\rho_{B0} + \ln \frac{\rho_{B0}}{\rho_{A0}}\Big]\right\}d\mathbf{r} \quad (5)$$

as in [29], where $\int \phi_1 C^{(2)} * \phi_2 d\mathbf{r} = \int \phi_2 C^{(2)} * \phi_1 d\mathbf{r}$ is written as $\int \phi_1 C^{(2)} \phi_2 d\mathbf{r}$ for brevity. For an $AB$ compound, using the following substitutions, [30]

1. $\psi = 2c - 1 = \frac{\rho_A - \rho_B}{\rho_A + \rho_B}$
2. $n = \frac{\rho - \rho_0}{\rho_0}$
3. $\Delta C = \frac{\rho_0}{4}(C_{AA}^{(2)} + C_{BB}^{(2)} - 2C_{AB}^{(2)})$
4. $\delta C = \frac{\rho_0}{4}(C_{AA}^{(2)} - C_{BB}^{(2)})$
5. $\overline{C} = \frac{\rho_0}{4}(C_{AA}^{(2)} + C_{BB}^{(2)} + 2C_{AB}^{(2)})$

expanding $n$ and $\psi$ to fourth order, and ignoring linear terms as is customary [5] results in

$$F = kT\rho_0 \int_V \left\{ \frac{n}{2}\Big[1 - (\overline{C} + 2\delta C \psi + \psi \Delta C \psi)\Big]n - \frac{1}{6}n^3 + \frac{1}{12}n^4 + \frac{1}{2}\psi\Big[\ln \frac{\rho_{B0}}{\rho_{A0}} + \frac{2\Delta C}{\rho_0}(\rho_{A0} - \rho_{B0})\Big](n+1) + \frac{1}{2}\psi(1 - \Delta C)\psi + \frac{1}{12}\psi^4 - n\delta C\psi - n\psi\Delta C\psi + \frac{1}{2}n\psi^2 \right\}d\mathbf{r}. \quad (6)$$

As is typical in PFC models, $n$ is interpreted as the normalized atomic density and $\psi$ as the normalized difference in composition. Because the $\psi$ expansion is performed around $\psi = 0$, this model is only appropriate for systems where $c \simeq 1/2$. Of course, a more general model could be derived by expanding around a generic concentration, at the cost of increased complexity. In the random binary alloy case, $\psi$ is assumed to vary on a length scale much larger than the atomic unit cell. In this limit, Eq. 6 reduces to the free energy of the binary alloy [29]. However, drawing inspiration from the theory of concentration waves, our model regards $\psi$ as a field that specifies the chemical identity of atoms inside a unit cell [31].

Two additional simplifying assumptions are now made in order to make Eq. 6 more tractable. First, the fourth-order convolution term, $n\psi\Delta Cn\psi$, is neglected for numerical ease. Second, it is assumed that $C_{AA}^{(2)} = C_{BB}^{(2)}$ and $\rho_{A0} = \rho_{B0}$. This assumption treats pure A and B as equivalent and yields a phase diagram that is symmetric

about a 50-50 stoichiometry. Using these two assumptions, Eq. 6 simplifies to

$$F = kT\rho_0 \int_V \left[ \frac{1}{2}n^2 - \frac{1}{6}n^3 + \frac{1}{12}n^4 - \frac{1}{2}n\overline{C}n + \frac{1}{2}\psi^2 + \frac{1}{12}\psi^4 - \frac{1}{2}\psi\Delta C\psi + \frac{1}{2}n\psi^2 - n\psi\Delta C\psi \right] d\mathbf{r}. \quad (7)$$

In a situation where the effect of the ordering on the free energy is considered minor in comparison to the overall crystal structure, we phenomenologically add a factor of $\epsilon$ to all terms that involve ordering,

$$\tilde{F} = \int_V \left[ \frac{1}{2}n^2 - \frac{1}{6}n^3 + \frac{1}{12}n^4 - \frac{1}{2}n\overline{C}n + \epsilon\left(\frac{1}{2}\psi^2 + \frac{1}{12}\psi^4 - \frac{1}{2}\psi\Delta C\psi + \frac{1}{2}n\psi^2 - n\psi\Delta C\psi\right) \right] d\mathbf{r} \quad (8)$$

where $\tilde{F} = F/(kT\rho_0)$. This $\epsilon$ factor will be further explained in sec. III.

### B. Correlation Function Determination

In order to construct $\Delta C$ and $\overline{C}$, we adapt a methodology similar to Greenwood et al. by considering their form in Fourier space [10]. In this section, we demonstrate a method to determine where the peaks of the correlation function in Fourier space should be for a given target structure.

In the case of a bulk crystal, we can exactly express the density as a Fourier series over all the reciprocal lattice vectors, namely

$$\rho_A = \bar{\rho}_A + \sum_{\mathbf{k}} A_A(\mathbf{k}) e^{i\mathbf{k}\cdot\mathbf{r}}. \quad (9)$$

In the limit where $\rho_A$ consists of delta functions weighted by $f_A$ at each atomic position, then orthogonality gives easily calculable values for the amplitudes in terms of structure factors, namely

$$A_A(\mathbf{k}) = \frac{f_A}{V} \sum_{j\in\text{cell}} e^{-i\mathbf{k}\cdot\mathbf{r}_j} = \frac{1}{V} S_A(\mathbf{k}) \quad (10)$$

where $V$ is the volume of the unit cell, $j$ indexes through all atoms in the unit cell, and $S_A \equiv f_A \sum e^{-i\mathbf{k}\cdot\mathbf{r}_j}$ is the structure factor of $A$, consistent with the definition by Cullity and Kittel [32, 33] (this definition is not universal [34, 35]). Fourier expanding $n$ and $\psi$ using the same reciprocal lattice vectors as A and B results in

$$A_n = \frac{1}{\rho_0}(A_A + A_B) \quad (11)$$

and

$$A_\psi \simeq \frac{1}{\rho_0}(A_A - A_B), \quad (12)$$

where the approximation $\rho_A(\mathbf{r}) + \rho_B(\mathbf{r}) \simeq \rho_0$ (equivalent to assuming $n(\mathbf{r})$ is small) is used for deriving the latter expression.

Because the exact values calculated from Eq. 11 and Eq. 12 depend on the delta function assumption, these values will never occur in the numerical model. What is more important than the exact value calculated is whether the amplitude for a given $\mathbf{k}$ value is zero or nonzero, since the free energy is only affected by the value of the correlation function at $\mathbf{k}$ values when the amplitude for that same $\mathbf{k}$ is nonzero. This is because after Fourier expanding $n$,

$$\int_V n\overline{C}n d\mathbf{r} = V \sum_{\mathbf{k}} |A_n(\mathbf{k})|^2 \hat{\overline{C}}(|\mathbf{k}|), \quad (13)$$

where the hat denotes the Fourier transform. Since the amplitude for each individual component is proportional to its structure factor, we define $S_n \equiv S_A + S_B$ and $S_\psi \equiv S_A - S_B$, and we expect the underlying symmetries of these "structure factors" to preserve the symmetries in the amplitudes as is the case for experimental structure factors [32]. In order to calculate $S_n$ and $S_\psi$, the simplifying assumption that $f_A = f_B = f$ is employed, consistent with assuming $\rho_{A0} = \rho_{B0}$.

As an example, let us calculate the peak locations for the B2 system explicitly (Fig. 1). Practically, the relation $\Delta\phi(\mathbf{k}) = \mathbf{k} \cdot \mathbf{x}_0$ is helpful, where $\Delta\phi(\mathbf{k})$ is the change in phase in the structure factor as a result of moving the origin of the coordinate system by $\mathbf{x}_0$. Denoting $\mathbf{k} = h\mathbf{b}_1 + k\mathbf{b}_2 + l\mathbf{b}_3$, where $b_i$ is the $i$th reciprocal lattice vector, then

$$S_A = f \text{ for all h, k, l} \quad (14)$$

$$S_B = \begin{cases} f & \text{if } h+k+l = 2m \\ -f & \text{if } h+k+l = 2m+1 \end{cases}$$

and

$$\begin{aligned} S_n &= 2f \text{ if } h+k+l = 2m \text{ and} \\ S_\psi &= 2f \text{ if } h+k+l = 2m+1 \end{aligned} \quad (15)$$

where $m$ is an integer. Thus, the only $\mathbf{k}$ that are non-zero are

$$n : \tfrac{2\pi}{a}\{110\}, \tfrac{2\pi}{a}\{200\}, \tfrac{2\pi}{a}\{211\}, ... \quad (16)$$

$$\psi : \tfrac{2\pi}{a}\{100\}, \tfrac{2\pi}{a}\{111\}, \tfrac{2\pi}{a}\{210\}, ... \quad (17)$$

where {} denotes a family of reciprocal lattice vectors created by the permutation of the internal elements (for example, {110} includes 12 vectors: $\hat{x}+\hat{y}$, $\hat{x}-\hat{y}$, $\hat{x}+\hat{z}$, ...). The structure factor (Eq. 15) for $n$ is consistent with the missing reflections for a BCC lattice. Because the first nonzero $\mathbf{k}$ for $n$ has magnitude $2\pi\sqrt{2}/a$, and the first nonzero $\mathbf{k}$ for $\psi$ has magnitude $2\pi/a$, the ratio of the first peak in $\overline{C}$ in reciprocal space to the first in $\Delta C$ must be $\sqrt{2}$. This same procedure can be easily performed to construct correlation functions for other 50-50 stoichiometry compounds, with the results for the locations of the first three peaks for $n$ and $\psi$ for various compounds listed in Table I. All compounds listed in Table I can produce (at least) metastable structures using at most three peaks. However, many, such as B2, can exist with fewer, with the exact stability regimes depending on parameter choices. The model can be extended to other compounds without 50-50 stoichiometry through this same method, but as explained in sec. II A, rigorously the free energy should also be derived with that additional complication — an extension left for future work. Although the values of the correlation function at specific $|\mathbf{k}|$ values are the only quantities that determine equilibrium structure, the exact functional form of the correlation function is important for properties such as elastic constants, defect structures, and dynamics.

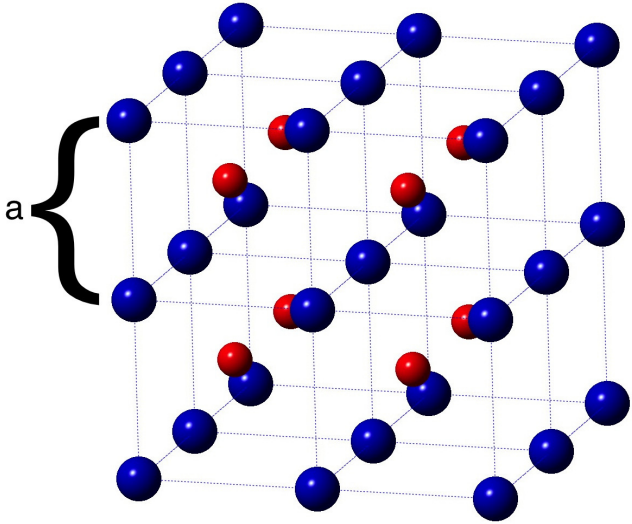

FIG. 1. Structure for a B2 crystal. The origin used for calculating the structure factor is at the center of a blue atom.

### C. XPFC Model

In this study, the XPFC formalism was employed [9, 10]. In the case of a single peak for $n$ and $\psi$,

$$\hat{\overline{C}} = B_x e^{-T/T_n} e^{-\frac{(k-k_n)^2}{2\alpha_n^2}} \text{ and } \Delta \hat{C} = D_x e^{-T/T_\psi} e^{-\frac{(k-k_\psi)^2}{2\alpha_\psi^2}},$$
(18)

where $k \equiv |\mathbf{k}|$, $B_x$ and $D_x$ are phenomenological constants, $T$ is a temperature-like parameter, $T_n$ controls the solid-liquid transition temperature, and $T_\psi$ controls the ordering temperature. The one-peak XPFC model has the advantage compared to polynomial PFC models in that the former does not exhibit significant changes in the equilibrium lattice constant with changes in average density or temperature, simplifying phase diagram construction. In polynomial PFC models, such as Eq. 1, the equilibrium lattice constant is determined by a competition between the energies of the primary frequency and the higher order harmonics; the primary frequency energy is minimized by a lattice constant that corresponds to the extremum of the correlation function while the higher frequencies decrease in energy by increasing the lattice constant [36]. However, the correlation function in XPFC quickly goes to zero for large $k$. Consequently, changing the lattice constant only negligibly affects the energetic contributions of higher frequency modes, and so the lattice constant is chosen purely so that the primary frequency of the structure coincides with the correlation function's maximum. Because the peak location in the correlation function is independent of temperature and composition, the lattice constant is not affected by these parameters.

This particular computational advantage of XPFC comes at the price of a less accurate one mode approximation however. Because polynomial correlation functions diverge for large $k$, their correlation functions strongly penalize higher order harmonics, resulting in free energies dominated by only the primary frequency [5, 12, 37]. In contrast, the XPFC correlation function is 0 for large $k$, rather than negative, so higher order harmonics play a larger role.

Unlike that shown in Eq. 18, previous XPFC models have assumed that the temperature is proportional to $\sigma$, using a temperature factor $\exp(\sigma^2/\sigma_M^2)$, equivalent to treating $\sqrt{T}$ as the temperature in Eq. 18 [9, 10, 15, 16, 38]. However, this relation is inconsistent with the usual temperature dependence of the Debye-Waller factor observed in diffraction experiments. When temperatures are much higher than the Debye temperature of the crystal, which is the case for PFC models [32], the atomic structure factor can be approximated as [33]

$$f = f_0 e^{-T/T_0} .$$
(19)

Assuming $\hat{C}(k) \propto f^\nu$, meaning the correlation function peak height scales with the atomic structure factor to some power, then in the case $T \ll T_0$, $\hat{C}(k)$ decreases linearly with $T$. Note that this is exactly the temperature dependence of the correlation function in PFC models with polynomial correlation functions and is consistent with the linear temperature dependence of the quadratic term in Landau models. For example, in Eq. 1, $\varepsilon$ is considered the variable proportional to the temperature, and it decreases the effective correlation function linearly. Because the original XPFC model goes as $\exp(\sigma^2/\sigma_M^2)$, in the limit $\sigma \ll \sigma_M$ the correlation function decreases quadratically with $\sigma$. Although understanding that in the original XPFC terminology $\sigma^2$ is the temperature parameter rather than $\sigma$ does not matter when fitting data at a particular temperature, nor does it matter when calculating the shape of phase diagrams qualitatively, it does affect the values of critical exponents (see sec. IV).

TABLE I. Table of peak locations, $|\mathbf{k}|_i \equiv k_i$, for the $n$ and $\psi$ correlation functions in reciprocal space for various crystal structures. The $k_i$ values for each structure were calculated using a square/cubic unit cell with side lengths of $2\pi$, except p3m1 whose calculation was performed using a rectangular unit cell with dimensions $2\pi \times 2\sqrt{3}\pi$. Because the calculations are performed on nonprimitive unit cells for convenience, the total atoms per cell are listed in order to uniquely identify the unit cell geometry.

| Space Group | Example | Atoms/Cell | $k_n$ | $k_\psi$ |
|---|---|---|---|---|
| p4gm | {100} plane of NaCl | 4 | $2, \sqrt{8}, 4$ | $\sqrt{2}, \sqrt{10}, \sqrt{18}$ |
| p3m1 | 2d h-BN | 4 | $2/\sqrt{3}, 2, 4/\sqrt{3}$ | $2/\sqrt{3}, 4/\sqrt{3}, \sqrt{28/3}$ |
| $Pm\bar{3}m$ (B2) | CuZn | 2 | $\sqrt{2}, 2, \sqrt{6}$ | $1, \sqrt{3}, \sqrt{5}$ |
| $Fd\bar{3}m$ (B32) | NaTl | 16 | $\sqrt{8}, 4, \sqrt{24}$ | $\sqrt{3}, \sqrt{11}, \sqrt{19}$ |
| P4/mmm (L1$_0$) | CuAu | 4 | $\sqrt{3}, 2, \sqrt{8}$ | $1, \sqrt{2}, \sqrt{5}$ |

## III. B2 ORDERING

In the remainder of this paper, three B2 models based on Eq. 8 will be examined. The first model considers the limit $\epsilon \to 0$ in Eq. 8. Namely, the density field is considered completely independent of the composition field, but the composition field is dependent on the density field. This assumption is analytically equivalent to separating the free energy into two separate equations,

$$\mathcal{F}_n = \int_V \left[ \frac{1}{2}n^2 - \frac{1}{6}n^3 + \frac{1}{12}n^4 - \frac{1}{2}n\overline{C}n \right] d\mathbf{r} \quad (20)$$

$$\text{and } \mathcal{F}_\psi = \int_V \left[ \frac{1}{2}\psi^2 + \frac{1}{12}\psi^4 - \frac{1}{2}\psi\Delta C\psi + \frac{1}{2}n\psi^2 - n\psi\Delta C\psi \right] d\mathbf{r}$$

with $\tilde{F} = \mathcal{F}_n + \epsilon \mathcal{F}_\psi$. For dynamical behavior, evolution follows the typical simplified conserved Cahn-Hilliard equations [5]

$$\frac{\partial n}{\partial t} = M_n \nabla^2 \frac{\delta \mathcal{F}_n}{\delta n} \quad (21)$$

$$\frac{\partial \psi}{\partial t} = M_\psi \nabla^2 \frac{\delta \mathcal{F}_\psi}{\delta \psi}. \quad (22)$$

This approach, which we will call the "uncoupled" case, is both computationally cheaper compared to the "coupled" case (i.e. nonzero $\epsilon$) and is also significantly simpler with respect to phase diagram construction because the $n$ field in isolation has already been described by the original XPFC papers [9, 10].

Similar to the uncoupled model is the "weakly coupled" model, which uses a finite $\epsilon \ll 1$. For simplicity, the evolution equations for the coupled case simply replaces $\mathcal{F}_n$ and $\mathcal{F}_\psi$ with $\tilde{F}$ rather than the rigorously derived evolution equations described by Jugdutt [39] although both methods result in the same equilibrium states. Assuming $\epsilon \ll 1$ is physically reasonable because the energy associated with order-disorder transitions is significantly less than that for atomic rearrangements. For example, in the FeAl system at 0 K, the free energies of ordered B2 and disordered A2 structures are within 0.001 eV/atom of each other [40, 41]. In contrast, $B_h$, the FeAl ordered structure with the next lowest energy, is 0.125 eV/atom higher in energy [40]. Both the weakly coupled and uncoupled models result in phase diagrams with a line of second-order transitions.

The last model, the "strongly coupled" case, considers when $\epsilon = 1$, or equivalently Eq. 7. In this case, there are very different $n$ fields in the B2 versus A2 (disordered) phases, giving rise to first-order transitions. Further, because this model strongly couples the $n$ and $\psi$ fields, a disordered hexagonal rod phase is in competition with the B2 and A2 phases. For all parameter regimes tested with a single peak, B2-hexagonal rod coexistence occurred rather than B2-A2 coexistence. Consequently, an additional $k = 0$ peak was added to $\Delta C$ in order to preferentially stabilize the A2 phase over the hexagonal rod phase. Namely,

$$\Delta \hat{C} = D_0 e^{-\frac{k^2}{2\alpha_\psi^2}} + D_x e^{-T/T_\psi} e^{-\frac{(k-k_\psi)^2}{2\alpha_\psi^2}}, \quad (23)$$

where $D_0$ was phenomenologically chosen.

### A. Analytic Phase Diagram

Because the structure factors for $n$ and $\psi$ for the B2 system (Eq. 15) are real and only dependent on $|\mathbf{k}|$, a one mode approximation for $n$ and $\psi$ is

$$n = \bar{n} + A_n \sum_{j \in \{110\}} e^{i\mathbf{k}_j \cdot \mathbf{r}} \text{ and } \psi = \bar{\psi} + A_\psi \sum_{j \in \{100\}} e^{i\mathbf{k}_j \cdot \mathbf{r}}, \quad (24)$$

where, $\vec{k}_j = \frac{2\pi}{a}(h,k,l)$, $A_n$ and $A_\psi$ are constant real numbers, and $A_\psi$ is the system's order parameter. Substituting equations 18 and 24 into Eq. 20 and integrating with the equilibrium lattice parameter results in

$$\mathcal{F}_n = 6\bar{n}^2 A_n^2 + 16\bar{n}A_n^3 - 6\bar{n}A_n^2 + \frac{\bar{n}^4}{12} - \frac{\bar{n}^3}{6} + \frac{\bar{n}^2}{2} + 6A_n^2(1 - B_x e^{-\frac{T}{T_n}}) + 45A_n^4 - 8A_n^3, \quad (25)$$

$$\mathcal{F}_\psi = -6\bar{n}A_\psi^2 D_x e^{-\frac{T}{T_\psi}} + 3\bar{n}A_\psi^2 + 3\bar{\psi}^2 A_\psi^2 +$$
$$\frac{1}{2}\bar{n}\bar{\psi}^2(1-2D_0) + \frac{\bar{\psi}^4}{12} + \frac{\bar{\psi}^2}{2} + A_\psi^2\left[12A_n(1-2D_x e^{-\frac{T}{T_\psi}})\right.$$
$$\left. -3D_x e^{-\frac{T}{T_\psi}} + 3\right] + \frac{15A_\psi^4}{2} - \frac{D_0}{2}\bar{\psi}^2. \quad (26)$$

The phase diagram was constructed using four methods using the parameters shown in Table II, which was also used for the numerical phase diagram (Fig. 5 and 6). The first and simplest method used the uncoupled free energy. These calculations were performed at $\bar{n} = 0$ by both a common tangent construction and by solving $\partial^2 \mathcal{F}_\psi^{\text{order}}/\partial A_\psi^2 = 0$ (Fig. 2). Both techniques yielded the same result, implying a second-order transition. The second method was similar, except that it investigated the weakly coupled model with $\epsilon = 0.05$. Unsurprisingly, it gave a similar result. Although constructing the phase diagram at $\bar{n} = 0$ is a common approximation for a constant pressure phase diagram [14], it is known to not be thermodynamically consistent [39, 42]. For the third method, the phase boundary was calculated for the weakly coupled model using the true two-phase equilibrium conditions

$$\mu_{eq} = \left(\frac{1}{\bar{n}+1}\frac{\partial \bar{F}}{\partial \bar{\psi}}\right)\bigg|_{B2} = \left(\frac{1}{\bar{n}+1}\frac{\partial \bar{F}}{\partial \bar{\psi}}\right)\bigg|_{A2} \quad (27)$$

$$\left(\frac{\partial \bar{F}}{\partial \bar{n}} - \bar{\psi}\mu_{eq}\right)\bigg|_{B2} = \left(\frac{\partial \bar{F}}{\partial \bar{n}} - \bar{\psi}\mu_{eq}\right)\bigg|_{A2} \quad (28)$$

$$p = \left((\bar{n}+1)\frac{\partial \bar{F}}{\partial \bar{n}} - \bar{F}\right)\bigg|_{B2} = \left((\bar{n}+1)\frac{\partial \bar{F}}{\partial \bar{n}} - \bar{F}\right)\bigg|_{A2} \quad (29)$$

where $\bar{n}$ and $\bar{\psi}$ can differ between the two phases and $\bar{F} \equiv \tilde{F}/V$ [42]. Notice that using these equilibrium conditions, the densities of the two phases can differ. However, for the weakly coupled model, the solution within numerical resolution is coexistence at the same densities and compositions, implying a second-order transition. For a judicious choice of pressure ($p = -0.063$), $\bar{n} \simeq 0$ at all temperatures, and the phase boundary is very similar to those calculated by the previous techniques.

The last phase diagram construction was done for the strong coupling case with $p = -0.063$, $\epsilon = 1$, and $D_0 = 0.5$ (Fig. 3). Unlike weak coupling, this results in first-order transitions with two-phase coexistence. Calculating the order parameter of the B2 phase for increasing temperatures at $\bar{\psi} = 0$, where $\bar{n}$ is the B2 coexistence value, results in a discontinuity in the order parameter at the point where the two phase boundaries meet (Fig 4). Consequently, this point is a congruent point. In contrast, the order parameter would go continuously to zero if it was a critical point with a second-order transition.

In order to use Eq. 25 and 26 to describe a real compound such as FeCo [43], one can simply match properties such as the temperature and second derivative of the phase boundary at the critical point measured from an experimental phase diagram to the corresponding values at the critical point in the analytic (or numeric) model.

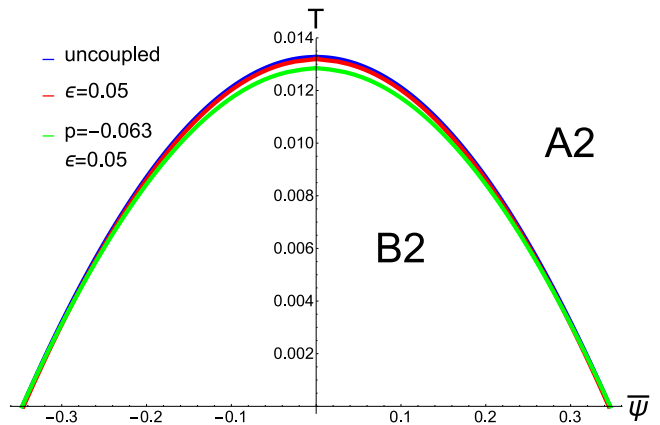

FIG. 2. Phase diagram for analytical model. The uncoupled and coupled models using a common tangent equilibrium condition are compared to a constant pressure phase diagram. As all phase diagrams are very similar, including constant pressure and weak coupling complications appear to have little effect.

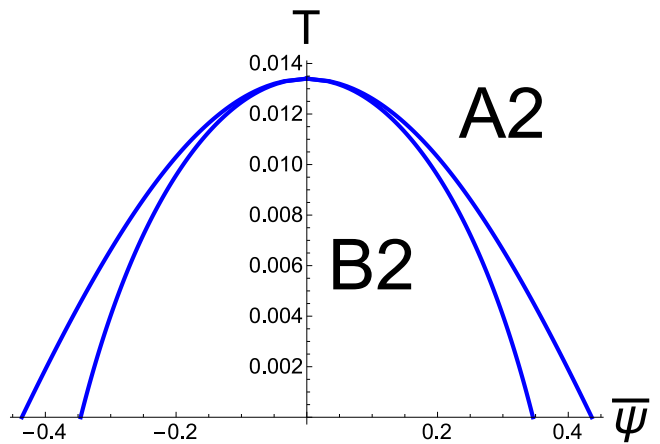

FIG. 3. Phase diagram for analytical model in strongly coupled case, $\epsilon = 1$, at pressure = -0.063.

### B. Elasticity

In order to use the traditional elastic constant calculation procedure and not consider the effects of hydrostatic pressures [44], only the case where $\bar{n} = \bar{\psi} = 0$ is considered. Since $\rho_A$ and $\rho_B$ ought to undergo identical strains, the $\psi$ field is strained identically to the $n$ field. Substituting in Eq. 24 into Eq. 8 and applying isotropic, biaxial, and simple shear strains as described by Pisutha-Arnond [44] results in the elastic constants

$$C_{11} = \frac{A_n^2 B_x e^{-\frac{T}{T_n}}}{\alpha_n^2} + \frac{\epsilon(8A_n+1)A_\psi^2 D_x e^{-\frac{T}{T_\psi}}}{2\alpha_1^2} \quad (30)$$

$$C_{12} = C_{44} = \frac{A_n^2 B_x e^{-\frac{T}{T_n}}}{2\alpha_n^2}. \quad (31)$$

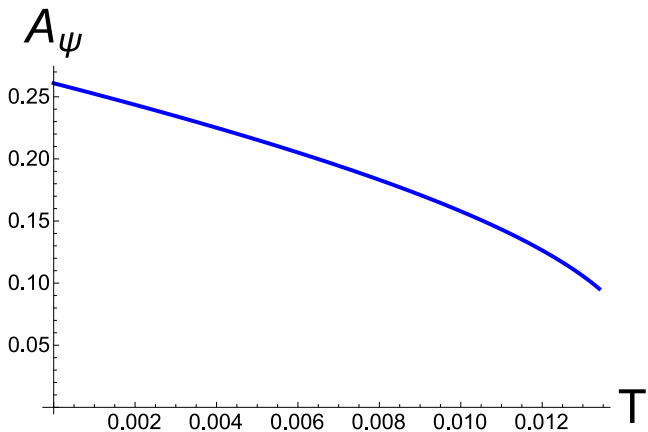

FIG. 4. Plot of the order parameter as a function of $T$ for $\bar{\psi} = 0$. Because of the discontinuity of the order parameter from nonzero to zero, the transition is first-order with a congruent point at $T \simeq 0.0134$.

Note that the strains only couple to the correlation terms in the free energy and are not affected by the ideal entropy of mixing terms. Spatschek and Karma ([45], Eq. 59) make two related predictions for a free energy consisting of a single peak in the correlation function. First, they predict that

$$\frac{C_{11}}{2} = C_{12} = C_{44} = -\frac{1}{2}C''(q_0)q_0^2 A_{110}^2 , \qquad (32)$$

which, using Eq. 16, agrees with Eq. 30 and 31 in the limit $\epsilon = 0$. Second, they predict that $C_{12}$ and $C_{44}$ depend on neither {100} nor {200} amplitudes. This explains why only $C_{11}$ depends on the ordering amplitude, since the ordering wave is a {100} mode.

Another way to help rationalize the lack of ordering dependence of the $C_{44}$ elastic constant is that $C_{44}$ physically represents shear on {100} planes. These planes are of a single atom type, however, and thus not really dependent on ordering (to lowest order). The shear mode on the {110} planes is proportional to $C_{11} - C_{12}$, and as expected does depend on $A_\psi$.

Equations 30 and 31 also agree with the Monte Carlo simulations of a B2 compound by Castán and Planes who found that the shear modulus, $(C_{11} - C_{12})/2$, is linearly proportional to the long-range order parameter squared [46]. However, eqs. 30 and 31 are inconsistent with experimental data for CuZn as McManus found no anomalous behavior near the critical temperature for $C_{11}$, but an abrupt change in $\frac{dC_{44}}{dT}$ shortly before the critical temperature [47].

### C. Numerical Phase Diagram

To confirm the behavior of the one-mode model phase diagram in the case where all frequencies were included, the phase diagram was calculated numerically at $\bar{n} = 0$ for the uncoupled (Fig. 5), weakly coupled ($\epsilon = 0.05$, Fig. 5), and strongly coupled ($\epsilon = 1, D_0 = 0.5$) cases (Fig. 6). Because the constant pressure condition was unimportant when constructing the analytic phase diagrams, only the $\bar{n} = 0$ method was employed for ease when constructing the numerical phase diagrams. Numerical construction of the phase diagrams validated the qualitative behavior of a curve of second-order and first-order transitions seen previously. Quantitative disagreement with analytical results are explained by the fact that the XPFC model permits high frequency modes, so the one mode approximation provides poor quantitative estimates of free energies (see sec. II C).

The free energies for the phase diagram were calculated by minimizing the free energy in a $2 \times 2 \times 2$ set of unit cells using the parameters found in Table II. Each system was initialized by either a prior equilibrated structure for different parameters or by a single mode approximation. Each system was then evolved using Eq. 21 and Eq. 22 using the standard semi-implicit integral spectral method [29]. Conditions near the phase transition were tested carefully to ensure that the order of the transition was determined correctly. The amplitude of the {100} peak from the numerical Fourier transform of $\psi$ was treated as the order parameter in the numerical model (Fig. 7). In the case of first-order transitions, the coexistence region was determined by fitting a fourth-order polynomial through a set of free energies for B2 and A2, and then finding the convex hull. The phase transition in the simulations can be understood as follows: At $\bar{\psi} = 0$, the ordered state for $\psi$ consists of only the ordering modes ({100}, {111}, ...). For $\bar{\psi} \neq 0$, the disordering modes ({110}, {200}, ...) appear and gradually increase in magnitude while the ordered reflections diminish (see sec. II B). At the phase transitions, only the disordered modes remain.

The first-order transition was also confirmed by equilibrating a $2 \times 2 \times 128$ simulation box of B2-A2 using $\bar{\psi} = 0.265$ and $T = 0.0144$. As predicted, B2 and A2 were found to coexist at equilibrium. Although the two phases had different average densities as suspected, $\bar{n}_{B2} \simeq 0.052$ and $\bar{n}_{A2} \simeq -0.027$, both densities were still close to zero. In order to speed up this large calculation, simple conserved global dynamics were used for both the $n$ and $\psi$ fields in this calculation [39, 48], namely

$$\frac{\partial n}{\partial t} = -\frac{\delta \tilde{F}}{\delta n} + \frac{1}{V}\int_V \frac{\delta \tilde{F}}{\delta n} d\mathbf{r} \qquad (33)$$

$$\frac{\partial \psi}{\partial t} = -\frac{\delta \tilde{F}}{\delta \psi} + \frac{1}{V}\int_V \frac{\delta \tilde{F}}{\delta \psi} d\mathbf{r}. \qquad (34)$$

TABLE II. Table of parameter values for phase diagrams.

| Quantity | Value |
|---|---|
| $M_n$ | 1 |
| $M_\psi$ | 1 |
| $T_n$ | 1 |
| $T_\psi$ | .2 |
| $k_n$ | 1 |
| $k_\psi$ | $1/\sqrt{2}$ |
| $D_x$ | 0.8 |
| $D_0$ | 0.5 (strongly coupled) |
|  | 0 (otherwise) |
| $B_x$ | 1 |
| $\alpha_n$ | 0.08 |
| $\alpha_\psi$ | 0.08 |
| $a_0$ | $2\pi\sqrt{2}$ |
| Mesh | 16 / unit cell dimension |
| $\Delta t$ | 0.015 |

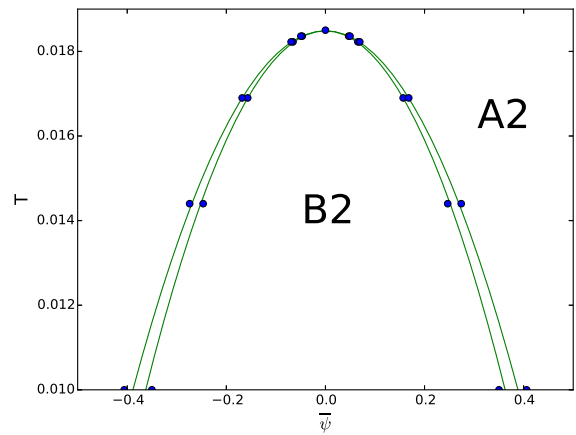

FIG. 6. Numerical phase diagram for the B2 system in the strongly coupled case. Consistent with the analytic results, the transition is always first-order, and the region between the phase boundaries is coexisting B2 and A2. The curves are parabolic fits through the numerically determined points.

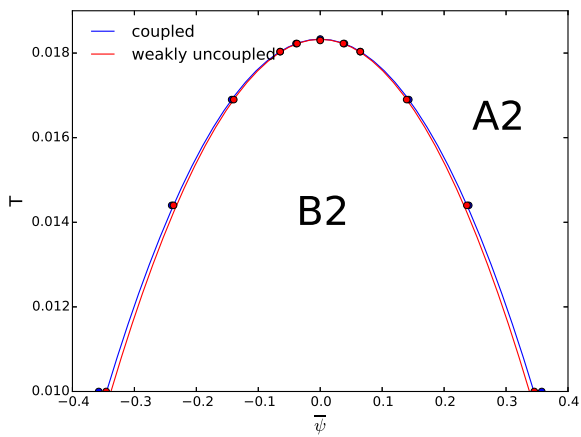

FIG. 5. Numerical phase diagram for the B2 system, with the uncoupled case (blue) and weakly coupled case (red). Consistent with the analytic results, the weakly coupled and uncoupled cases nearly overlap with second-order transitions. The curves are parabolic fits through the numerically determined points.

## IV. APPLICATIONS

### A. Antiphase Boundaries

As a test case for a system with a second-order transition, antiphase boundaries (APB) were investigated using the uncoupled model. Experimental evidence and Landau-Ginsburg theory predicts that the signed local order parameter, $\eta$, across the APB boundary is [26]

$$\eta(z;T) = A_\psi(T)\tanh\left(\frac{z-z_0}{2\xi(T)}\right) \quad (35)$$

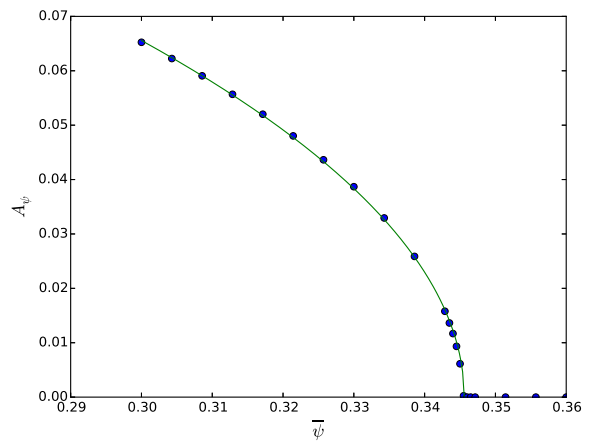

FIG. 7. Plot of the order parameter for the uncoupled case as a function of $\bar{\psi}$ for $T = 0.01$ using parameters found in Table II. The fit is to the function $A_\psi = A\sqrt{\bar{\psi}_c - \bar{\psi}}$, the function for a second-order mean-field transition as elaborated upon in section IV.

where $A_\psi$ is the unsigned bulk order parameter, $\xi$ is the correlation length, z is the coordinate perpendicular to the APB, and $z_0$ is the position of the interface. $A_\psi$, the APB energy $\gamma$, and $\xi$ all exhibit critical exponents. Meaning, near the critical point they are of the form $\sim \left(\frac{T_c-T}{T_c}\right)^{\nu_i}$, for some corresponding critical exponent $\nu_i$. This result was verified in the uncoupled B2 PFC model by initializing with a single mode approximation for two domains with opposite order parameters in a single grain, and the free energy was minimized using Eq. 21. The parameters in Table II were used except $\alpha_\psi = 0.25$ in order to reduce the width of the APB so that the simulation could be performed in a smaller domain. This change in $\alpha_\psi$ gave a new $T_c \simeq 0.01750$. A box size of $2 \times 2 \times 24$ unit cells was used with periodic boundary conditions (i.e. two identical APBs), and the results were confirmed in a few instances with a larger $2 \times 2 \times 96$ simulation domain. A comparison of a typical numerical result and a hyperbolic tangent profile fit is shown in Fig. 8. Because the local order parameter, $\eta$, is a function of position in this case, the amplitude of $\psi$ along a unit cell edge was used as a proxy for the order parameter. The measured critical exponent for $A_\psi$ was 0.51 with $R^2 \simeq 1.0 \times 10^{-6}$. The critical exponent for the APB energy was calculated using the typical form

$$\gamma = \frac{l_\perp}{2}(\bar{\mathcal{F}}_{\psi,\text{ APB}} - \bar{\mathcal{F}}_{\psi,\text{ eq}}), \qquad (36)$$

was 1.53 with $R^2 \simeq 2.7 \times 10^{-5}$ (Fig. 9), where $l_\perp$ is the length of the simulation box in the $z$ direction, $\bar{\mathcal{F}}_{\psi,\text{ APB}}$ is the free energy per volume measured with the APB, and $\bar{\mathcal{F}}_{\psi,\text{ eq}}$ is the bulk free energy per volume. Both of these exponents are consistent with the 1/2 and 3/2 exponents for $A_\psi$ and $\gamma$ respectively for an APB resulting from the simple Landau model

$$F = \int_V \left( \frac{r}{2}\eta^2 + \frac{u}{4}\eta^4 + \frac{K}{2}|\nabla\eta|^2 \right) d\mathbf{r}, \qquad (37)$$

where $r \propto (T_c - T)$ and $u$ and $K$ are independent of temperature. This is not surprising, as the PFC method is a mean field model and has thus have been shown to reduce to simple Landau models in appropriate limits [49]. Although comparison to this same Landau model would imply a critical exponent of -0.5 for $\xi$, unfortunately this exponent is difficult to determine. Because the critical exponent is highly sensitive to the method of fitting the $\psi$ profile, very small changes to the fitting methodology that only result in a few percent differences in the interface width give very different critical exponents. Because of a lack of a clear criteria for determining $\xi$, this critical exponent is left unreported.

As noted in section II, all these results are dependent on the new definition of temperature. Using the original XPFC temperature parameter, both the critical exponents change, deviating more from mean field theory, and $R^2$ increases.

### B. Grain Boundaries

As a further demonstration of the applicability of the this new method, ordering dynamics were investigated for symmetric tilt boundaries using the uncoupled model. The simulations were set up using the standard method for periodic grain boundaries [48]. The equilibrium density was determined by minimizing the energy with respect to $n$, and then initialized with Guassian noise for $\psi$ at T = 0.01 using the parameters in Table II except with $\alpha_\psi = 0.25$. The misorientation angle between the grains was 3.8°. The system first undergoes spinodal ordering. Then, domains grow and shrink in order to reduce the total APB energy. Interestingly, the model predicts that dislocation cores act as natural pinning points for APBs (Fig. 10). The dislocations pin the domain walls as excess disordered regions are created when a domain wall breaks free of the dislocation core.

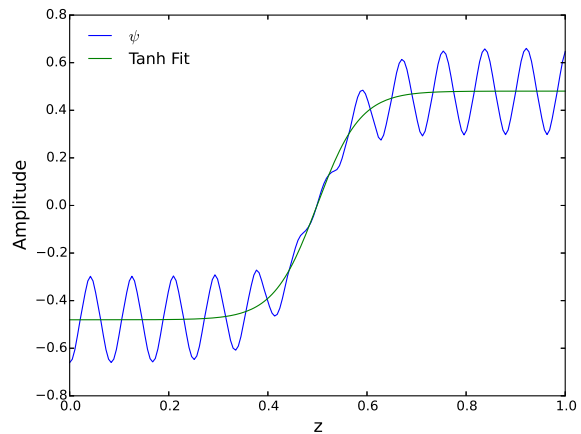

FIG. 8. Demonstration of a tanh profile across an APB. The $\psi$ amplitude along the edge of the unit cell (blue) and a fit to Eq. 35 (green) are plotted.

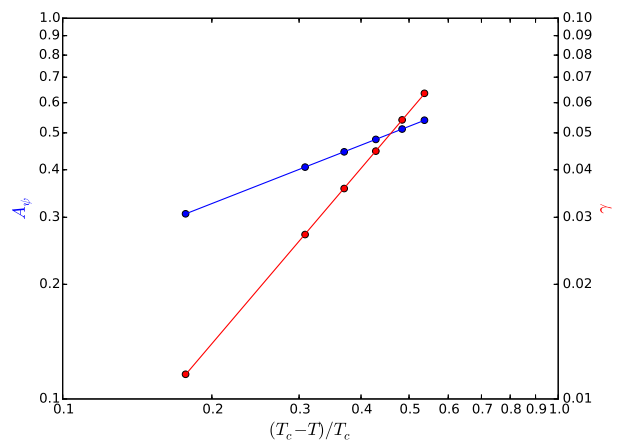

FIG. 9. Scaling behavior of bulk order parameter, $A_\psi$, and APB boundary energy, $\gamma$, for temperatures near $T_c$.

## V. SUMMARY

A two-component sublattice ordering model was derived that was shown to be capable of modeling a diverse set of ordered crystals. This model was then investigated in greater depth for the simple B2 system, with first-and second-order transitions found analytically and numerically. This model correctly predicts that not all the material's elastic constants are dependent on ordering. Finally, antiphase boundaries were explored in isolation and shown to reproduce standard mean-field results once the temperature parameter in XPFC was reinterpreted. In the context of large small-angle grain boundary simulations, these APBs were predicted to have their evolution pinned by dislocation cores. The model in the future can be used to study the dynamics of ordering in a wide range of crystals.

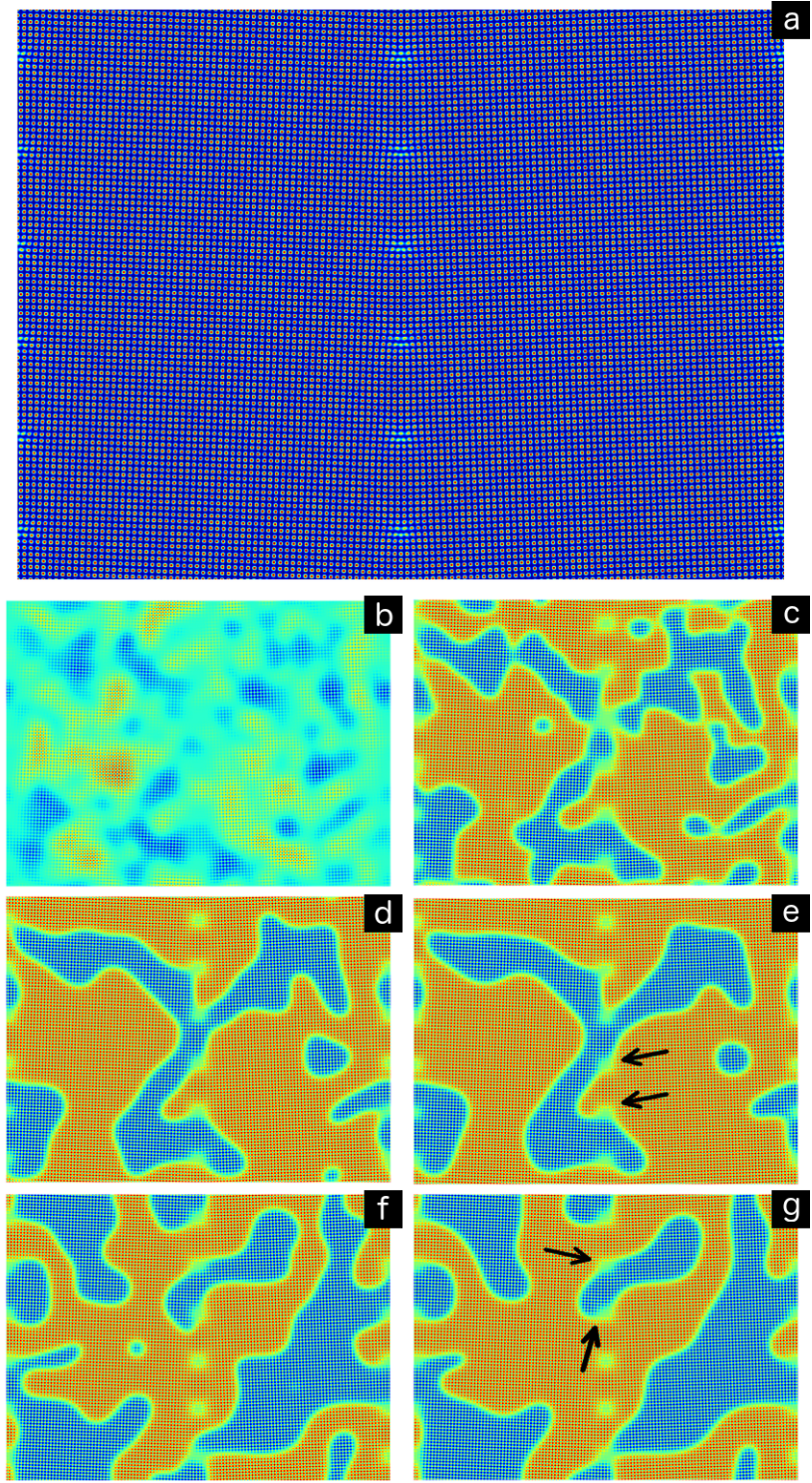

FIG. 10. Images for dynamic ordering of a B2 crystal, showing an xy slice through the middle z coordinate. (a) shows the static $n$ field for the simulations with 3.8° misorientation between the two grains. After initializing the $\psi$ field with Gaussian noise, (b)-(e) show snapshots of $\psi$ at progressively later stages in the evolution. (b) shows spinodal ordering while (c)-(e) show coarsening. The green disordered regions down the middle of (b)-(g) correspond to the edge dislocation cores in (a). Although much of the evolution can be understood through a simple reduction of mean curvature, the arrows point to examples of dislocation anchors that act to pin APB movement. The anchoring phenomenon was confirmed by initializing another simulation with the same conditions but a different set of random Gaussian noise to start, and two sample time steps from this second simulation are show in (f) and (g).

## ACKNOWLEDGMENTS

The author gratefully thanks Kevin McReynolds, Nana Ofori-Opuku, Zhifeng Huang, Matthew Peters, and the rest of the Voorhees Group for countless helpful discussions. Edwin Schwalbach is acknowledged for sharing his code repository. The National Science Foundation Graduate Research Fellowship, NSF Grant #1507033, and Northwestern University are thanked for financial support. K. R. E. acknowledges financial support from NSF Grant No. DMR-1506634.